# Characterizing the Dust Coma of Comet C/2012 S1 (ISON) at 4.15 AU from the Sun

Jian-Yang Li (李荐扬)[1], Michael S. P. Kelley[2], Matthew M. Knight[3], Tony L. Farnham[2], Harold A. Weaver[4], Michael F. A'Hearn[2], Max J. Mutchler[5], Ludmilla Kolokolova[2], Philippe Lamy[6], and Imre Toth[7]

[1] Planetary Science Institute, 1700 E. Ft. Lowell Rd., Suite 106, Tucson, AZ 85719, USA; jyli@psi.edu
[2] Department of Astronomy, University of Maryland, College Park, MD 20742, USA; msk@astro.umd.edu, farnham@astro.umd.edu, ma@astro.umd.edu, ludmilla@astro.umd.edu
[3] Lowell Observatory, 1400 W. Mars Hill Rd., Flagstaff, AZ 86001, USA; knight@lowell.edu
[4] Johns Hopkins University Applied Physics Laboratory, Space Department, 11100 Johns Hopkins Rd., Laurel, MD 20723, USA; Hal.Weaver@jhuapl.edu
[5] Space Telescope Science Institute, 3700 San Martin Drive, Baltimore, MD 21218-2463, USA; mutchler@stsci.edu
[6] Laboratoire d'Astrophysique de Marseille, UMR 7236, CNRS & Aix-Marseille Université, 38 rue Frédéric Joliot-Curie, 13388 Marseille Cedex 13, France; philippe.lamy@oamp.fr
[7] Konkoly Observatory, Research Centre for Astronomy and Earth Sciences, Hungarian Academy of Sciences, PO Box 67, 1525 Budapest, Hungary; tothi@konkoly.hu

**Abstract:**

We report results from broadband visible images of comet C/2012 S1 (ISON) obtained with the *Hubble Space Telescope* Wide Field Camera 3 on 2013 April 10. C/ISON's coma brightness follows a $1/\rho$ (where $\rho$ is the projected distance from the nucleus) profile out to 5000 km, consistent with a constant speed dust outflow model. The turnaround distance in the sunward direction suggests that the dust coma is composed of sub-micron-sized particles emitted at speeds of tens of meters s$^{-1}$. $A(\theta)f\rho$, which is commonly used to characterize the dust production rate, was 1340 and 1240 cm in the F606W and F438W filters, respectively, in apertures <1.6" in radius. The dust colors are slightly redder than solar, with a slope of 5.0±0.2% per 100 nm, increasing to >10% per 100 nm 10,000 km down the tail. The colors are similar to those of comet C/1995 O1 (Hale-Bopp) and other long-period comets, but somewhat bluer than typical values for short-period comets. The spatial color variations are also reminiscent of C/Hale-Bopp. A sunward jet is visible in enhanced images, curving to the north and then tailward in the outer coma. The 1.6"-long jet is centered at a position angle of 291º, with an opening angle of ~45º. The jet morphology remains unchanged over 19 hours of our observations, suggesting that it is near the rotational pole of the nucleus, and implying that the pole points to within 30º of (RA, Dec) = (330º, 0º). This pole orientation indicates a high obliquity of 50º-80º.

# 1. Introduction

The newly discovered Comet C/2012 S1 (ISON) has drawn intense interest since it was discovered in September 2012 (Nevski and Novichonok 2012). C/ISON is dynamically new (MPEC 2013-Q27), is in a sungrazing orbit, and was discovered more than a year before perihelion, creating a combination unique among the known comets. As a new comet visiting the inner solar system for the first time since being scattered to, and deeply frozen in, the Oort cloud (OC), C/ISON should preserve valuable information about the conditions in the solar nebula at the time of planetary formation. When C/ISON reaches perihelion, at a heliocentric distance of only 0.0125 AU (2.7 solar radii), its nucleus will experience temperatures so intense that even silicates and metals will vaporize, enabling observations of essentially its entire elemental composition. The discovery of C/ISON at 6.2 AU from the Sun offers an unprecedented opportunity to study the comet's behavior as it progresses from one thermal extreme to the other. The sungrazing perihelion will also make the comet an important probe for solar physics (e.g., Schrijver et al. 2012; Downs et al. 2013).

We observed C/ISON with the *Hubble Space Telescope* (*HST*) when the comet was 4.15 AU from the Sun, 4.24 AU from Earth, and at a phase angle of 13.7°. These observations occurred well before C/ISON crossed inside the "frost line" (2.5-3 AU) where water ice sublimation is expected to rapidly increase to become the dominant driver of activity (cf. Meech and Svoren 2004). *HST*'s high angular resolution provides the best view of C/ISON's inner coma prior to the onset of vigorous water-driven activity. Such imaging is rarely acquired for OC comets, and never before for sungrazing comets.

# 2. Observations and data reduction

We used *HST*'s Wide Field Camera 3 (WFC3) UVIS detector to image C/ISON through two broadband filters, F606W and F438W, over three separate orbits, from UTC 2013-Apr-10.2157 to 2013-Apr-10.9786. F606W is a "wide-*V*" filter with a pivot wavelength of 588.7 nm and a width of 218.2 nm (as defined in Dressel et al. 2012), and the F438W filter has a pivot wavelength of 432.5 nm and a width of 61.8 nm. The field-of-view (FOV) of the detector sub-frame was 40"×40" (1024×1024 pixels at 0.04" pixel$^{-1}$; 123 km pixel$^{-1}$).

All images were calibrated by the automated calibration pipeline at the Space Telescope Science Institute to remove the dark current and detector bias, and apply a flatfield correction (Rajan et al. 2010). We then used the *Astro-Drizzle* package to correct for the geometric distortions, rotate the image to align celestial north towards the top of the frame, and remove cosmic rays (Gonzaga et al. 2012). The comet showed a well-defined coma of ~3" (~9000 km) in the sunward direction, and a >30" long wide tail in the anti-sunward direction that extended beyond the FOV (Fig. 1). No changes in the comet were observed during the course of *HST* observations to the accuracy of the brightness measurements (~1%).

# 3. Analysis and Results
*3.1 Photometry*

After subtracting the sky background, determined from the sunward quadrant outside of the coma, we measured the total brightness of the comet in a series of circular apertures from 3 to 300 pixels in radius. The final total count rates were converted to flux and Vega magnitude using the calibration constants PHOTFLAM and PHOTZPT provided in the FITS file headers.

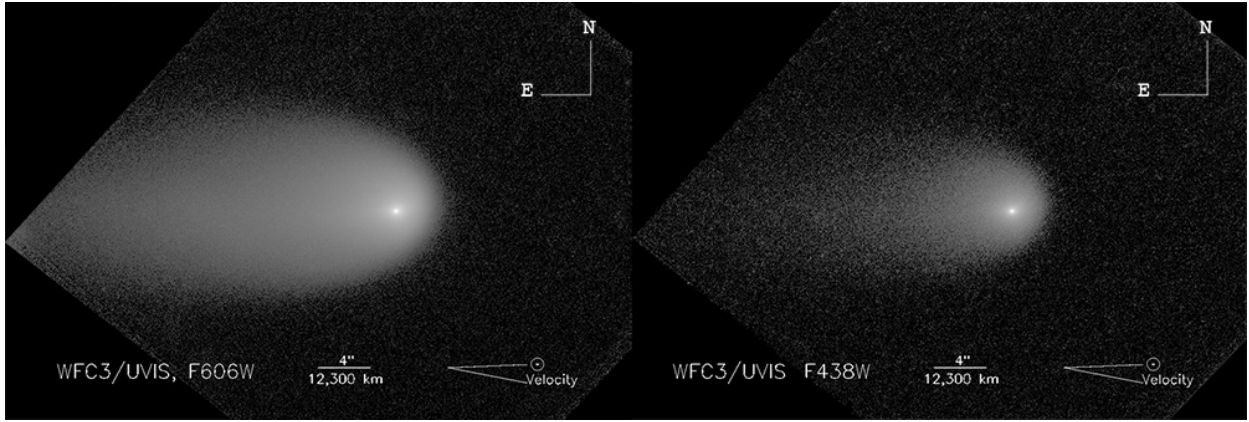

Fig. 1. *HST* images of Comet C/ISON, co-added from all exposures in the respective filters (F606W on the left, and F438W on the right). The brightness of the comet is displayed in logarithmic scale. Celestial north points up and east points to the left. The projected directions of the Sun and the comet's orbital velocity vectors are indicated.

The measurement uncertainties in our fluxes are <1%, excluding the systematic radiometric calibration uncertainty, which is estimated to be ~5% for the broadband filters we used, due to the unknown spectrum of the comet. To compare the comet's brightness with that of the Sun for the measurements of dust production rates and colors, we calculated the effective solar fluxes, $F_\odot$, and the apparent solar magnitudes, $M_\odot$, through each filter using a model solar spectrum (ASTM E-490 2006), $F_\odot(\lambda)$, and the corresponding system throughput curves, $T(\lambda)$,

$$F_\odot = \frac{\int F_\odot(\lambda) T(\lambda) d\lambda}{\int T(\lambda) d\lambda}$$

$$M_\odot = -2.5 \log\left(\frac{F_\odot}{PHOTFLAM}\right) + PHOTZPT$$

We derived $F_\odot(F606W)=1.73\times10^3$ Wm$^{-2}$μm$^{-1}$, $M_\odot(F606W)=-26.93$, $F_\odot(F438W)=1.80\times10^3$ Wm$^{-2}$μm$^{-1}$, and $M_\odot(F438W)=-26.08$. All the magnitudes quoted in this letter are calibrated to the Vega magnitude system (in which Vega has M=0 at all wavelengths).

The comet's total magnitude in a 1.6"-radius aperture is 17.13±0.05 in F606W and 18.08±0.05 in F438W. We did not observe any temporal brightness variations greater than the photometric uncertainty. This lack of photometric variability is consistent with observations by the *Deep Impact* flyby spacecraft in January 2013 (Farnham et al. 2013), and confirmed by the *Spitzer Space Telescope* observations in June 2013 (Lisse et al. 2013), although these measurements all had larger apertures. The lack of rotation-related brightness variation in the dust coma over almost half a year strongly suggests that any diurnal modulation of the dust activity at this time is less than a few percent of the total dust production and too weak to be detected.

The dust production rate in a 1.6"-radius aperture, quantified through the product $A(\theta)f\rho$, where $A(\theta)$ is the albedo of coma dust at phase angle θ, *f* is the filling factor of dust, and $\rho$ is the projected distance to the nucleus (A'Hearn et al. 1984), is 1340 cm in the F606W filter and 1240 cm in the F438W filter, with ~5% uncertainty for both values. We find that $A(\theta)f\rho$ remains nearly constant inside 1.6", but drops off for larger apertures due to the steeper than $1/\rho$ falloff beyond this point (§3.2). When similar apertures are used, our measurements are consistent with those reported by other observers (Schleicher 2013a, b; Farnham et al. 2013; Bodewits et al. 2013). Our *HST* measurements fall into the nearly constant dust activity trend of C/ISON from January to June (Meech et al. 2013).

*3.2 Coma radial profile*

The azimuthally averaged radial profile of the coma of C/ISON closely follows a $1/\rho$ surface brightness distribution in both filters for $\rho$ between ~0.1" (~300 km) and ~1.6" (~5000 km). The best-fit slopes for the azimuthally averaged brightness distribution are -1.0135±0.0002 and -1.024±0.002 for F606W and F438W, respectively. Outside of this region, the slope is steeper than $1/\rho$. The slightly steeper slope in F438W indicates the coma reddens with $\rho$ (§3.3). The $1/\rho$ brightness profile in the inner coma is consistent with a steady state dust outflow model, where dust leaves the nucleus isotropically at a constant speed, suggesting that there is active sublimation of volatiles driving the dust activity.

Based on the detection of CO radio emission in March 2013 (Biver, priv. comm.), and the upper limit of CO measured in May 2013 that is somehow lower than the detection in March (A'Hearn et al. 2013), some researchers (e.g., A'Hearn et al. 2013; Meech et al. 2013) have suggested that the stagnation observed in the comet's visual magnitude during January-May 2013 might be caused by a decrease in CO activity (following an earlier prolonged episode of high CO activity), leaving a residual, slowly dissipating dust coma that probably contains water ice grains. In this scenario, it is impossible to predict the brightness profile of the coma without more observational constraints on the dust properties and detailed numerical modeling, although the color variation observed in the coma (§3.3) appears consistent with this scenario. On the other hand, the *Spitzer* detection of a gas coma, which could be CO and/or $CO_2$ (Lisse et al. 2013), indicates that $CO_2$ might be the dominant driver to sustain the $1/\rho$ brightness profile.

*3.3 Color of the dust coma*

For a 1.6"-radius aperture, C/ISON has [F438W]-[F606W] = 0.95±0.03 mag, which can be compared to the corresponding solar color index of 0.86 mag. Thus, the overall dust color measured within our aperture is redder than the Sun by 0.09 mag, and the average red slope is 5.0±0.2% per 100 nm. This dust color is consistent with those of dynamically new comets, but is bluer than for most periodic comets (Jewitt and Meech 1986; Lowry et al. 1999; Hadamcik and Levasseur-Regourd 2009). The dust coma of C/ISON shows obvious color variations between 5000 km and 10,000 km from the nucleus, where the red slope increases from ~6% per 100 nm to ~12% per 100 nm down the tail (Fig. 2). A coma like this, becoming redder at larger distances is unusual. We compared the transmission profiles of the two wideband filters used and the typical visible light spectrum of comets (cf. Feldman et al. 2004) to determine the possible contribution of emission lines from C/ISON's gas coma. The observed flux in the F606W filter should have negligible gas contamination. For the F438W filter, the primary contaminant is the wide emission band from the $C_3$ coma, but the effect should be less than a few percent. Thus, our data suggest that the observed color variation is a real effect, possibly caused by changes in the composition or size distribution of the grains, as discussed further below. We also note that similar color variation was observed for the coma of Comet C/1995 O1 (Hale-Bopp), though at much larger spatial scales (Weiler et al. 2003).

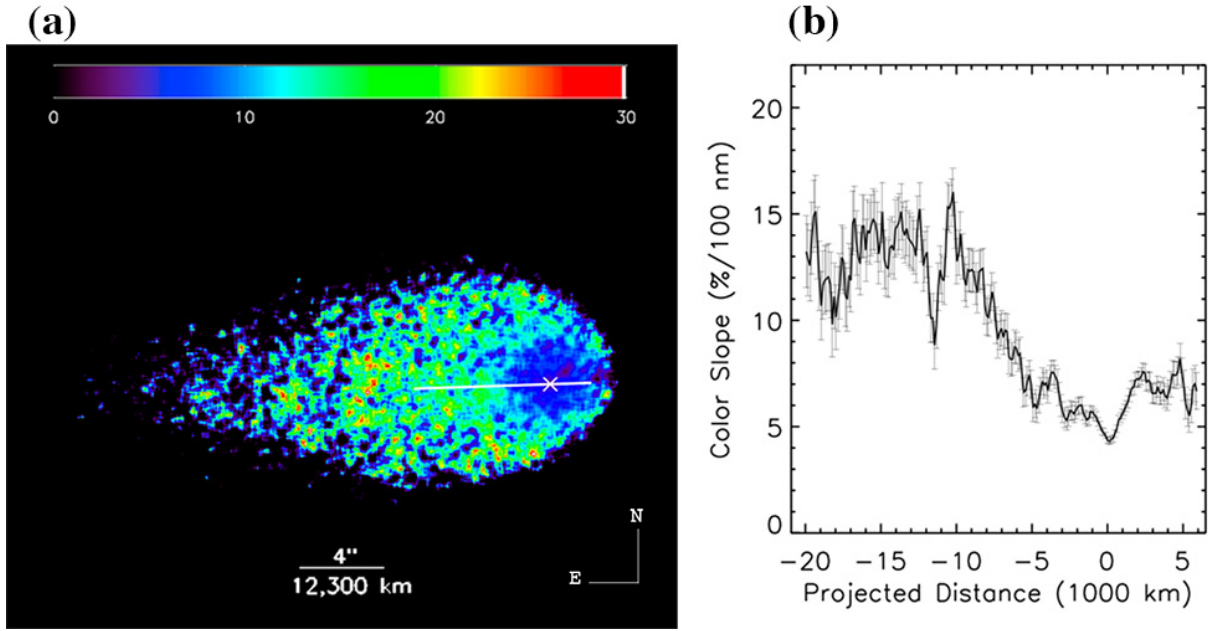

Fig. 2. Dust colors in the coma of C/ISON. Panel (a) shows the color slope (% per 100 nm) map, smoothed by a 9-pixel box moving average. The cross symbol marks the position of the nucleus. Panel (b) shows the color slope along the sunward-antisunward line marked in panel (a). Beyond 6000 km in the sunward direction, and beyond ~15,000 km in the antisunward direction, the F438W signal is weak and color measurements are not reliable.

Particle size sorting is a common phenomenon in comets, caused by terminal velocities (when dust is no longer being accelerated by gas drag) and/or radiation pressure efficiency having different influences for various sizes of grains. C/ISON's orbit is such that it is plunging straight toward the Sun and our observations had a low phase angle, so the coma and tail are highly projected. The effect of solar radiation pressure in the apparent segregation of particle sizes is negligible at the size of the observed coma. Therefore, if there is observable particle size sorting in the coma, then the different terminal velocities of large and small particles should dominate the process, resulting in relatively large, slow moving particles dominating the inner part of the coma and relatively small, fast moving particles dominating the outer part. This effect is well known and could dominate the size sorting process (Fink and Rubin 2012). If we assume the same composition for dust grains of all sizes, then larger grains should appear relatively redder than smaller grains (Kolokolova et al. 2001; Li and Greenberg 1997). Therefore, this dynamical argument suggests that the inner part of the coma should appear redder than the outer part, opposite of the measured behavior in C/ISON (Fig. 2). So particle size variation in the coma is not likely the cause of the color variations. Note that for large but fluffy aggregates, which are substantially more efficiently affected by both gas drag and solar radiation pressure than compact particles or aggregates (e.g., Nakamura et al. 1994; Nakamura and Hidaka 1998; Köhler et al. 2007), the terminal velocity could be close to that of small monomer particles. But it is unlikely for fluffy aggregates to reverse the size sorting. Detailed modeling will require the properties of the dust coma, such as particle ejection velocity, size distribution, etc. that are not currently available for this comet, and will be the subject of future studies.

On the other hand, Kolokolova et al. (2003) showed that change of water ice abundance could explain the observed color variations in the coma and jet of C/Hale-Bopp when it was at 3 AU from the Sun. It is conceivable that the coma of C/ISON could contain either water ice-rich aggregates and/or small, relatively pure water ice particles at 4.15 AU from the Sun. As those particles leave the nucleus, ice sublimates, causing the aggregates to become less ice-rich, and/or removing small ice particles. Because water ice appears blue or gray in visible wavelength, and small particles are relatively bluer than large particles, this scenario will lead to reddening with cometocentric distance in the coma. Abundant water ice-rich particles have been directly observed near the nucleus of 103P/Hartley 2 (A'Hearn et al. 2011; Kelley et al. 2013), and inferred for Comet C/2009 P1 (Garradd) (Paganini et al. 2012; Combi et al. 2013). Although those comets were observed at <2.5 AU from the Sun, it is known that water ice in Hartley 2 is associated with $CO_2$ outgassing (A'Hearn et al. 2011), suggesting that ice grains could also be driven out from C/ISON's nucleus and present in the coma. At 4 AU from the Sun, the lifetime of 1 μm sized dirty ice grains is several hours (Beer et al. 2006), and that of ice embedded in aggregates is likely weeks (Gustafson 1994). These lifetimes allow icy particles to travel $10^3$-$10^5$ km for typical grain velocities (10-100 m/s), consistent with the distance of the strongest color change we observed in C/ISON's coma. The water ice hypothesis does not necessarily conflict with the absence of ice absorption features in the 0.8-2.5 μm near-infrared spectrum obtained in mid-May (Yang 2013). It is well known that as the size of icy particles decreases, their 1.5 and 2.0 μm absorption bands become shallower and eventually unobservable as particles become sub-μm (e.g., Hansen and McCord 2004).

*3.4 Jet and rotational pole*

After removing the 1/ρ brightness distribution from the images of C/ISON, a prominent jet is visible in the sunward direction, and an enhancement of the coma brightness is visible in the anti-sunward direction in both filters (Fig. 3). The jet is located at a position angle 291°±3°, with an opening angle of ~45°, a projected length of ~1.6", and a slight curvature towards the north near the end. The anti-sunward feature is ~12° away from the opposite direction of the projected orbital velocity vector of the comet, most consistent with dust particles swept by solar radiation

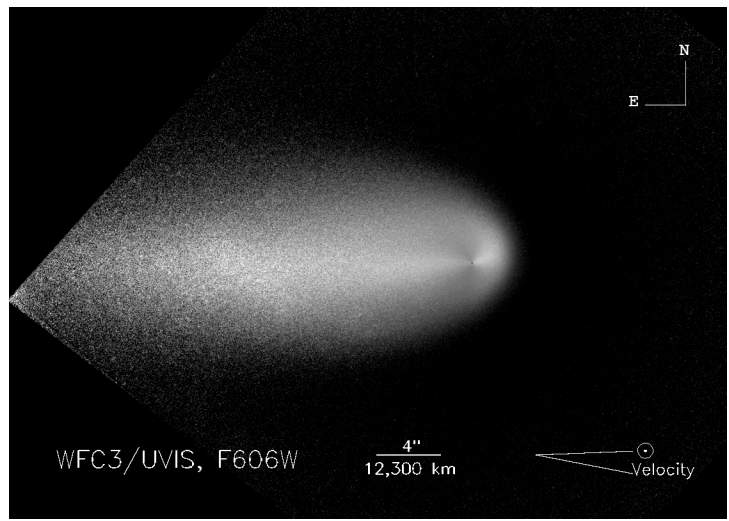

Fig. 3. The F606W image, after division of a 1/ρ image, showing the sunward jet, as well as the anti-sunward enhancement along the tail direction.

pressure towards the tail, rather than large dust grains forming a trail.

The jet morphology provides a means of constraining the orientation of the nucleus' spin axis. Analyses of the brightness and structure show no detectable changes in the jet morphology in any of our images. This jet has also been noticed in ground-based images, albeit at much lower spatial resolution, and remains relatively unchanged from March to May (Knight et al. 2013; Samarasinha, priv. comm.). The unchanging morphology of the jet suggests that the active area is located close to the pole, so that the outflowing dust sweeps out a narrow cone centered on the spin axis. In this scenario, the center line of the jet defines the spin axis as projected onto sky. In three dimensions, the axis lies along the plane formed by the linear jet and the line of sight. Extending this plane to its intersection on the celestial sphere produces the half-great circle shown in Fig. 4, which defines the family of possible solutions for the pole orientation.

As viewed from Earth, the observing aspect of C/ISON only changed by 7° from its discovery until the comet entered solar conjunction in mid-June. Having only a single observational geometry means that a unique pole solution cannot be derived. However, the appearance of the jet can be used to further reduce the range of acceptable rotation pole locations. The jet's well-defined nature suggests that it is not highly projected, likely lying at an angle >30° from the line of sight, and because it is active, the pole must point to within 90° of the Sun (and likely <75° to allow sufficient sunlight to generate activity). Satisfying these conditions indicates that the pole lies somewhere along the band stretching between RA, Dec = (310°, -15°) and (355°, +1 5°) (Fig 4). This result means that the obliquity lies in the range 100°-130° or 50°-80°, depending on whether the solution represents the positive or negative pole with respect to the direction of rotation. The positional uncertainty perpendicular to the line of solutions is from the 3º uncertainty in the measured PA of the jet, and is delineated by the shaded areas in the plot.

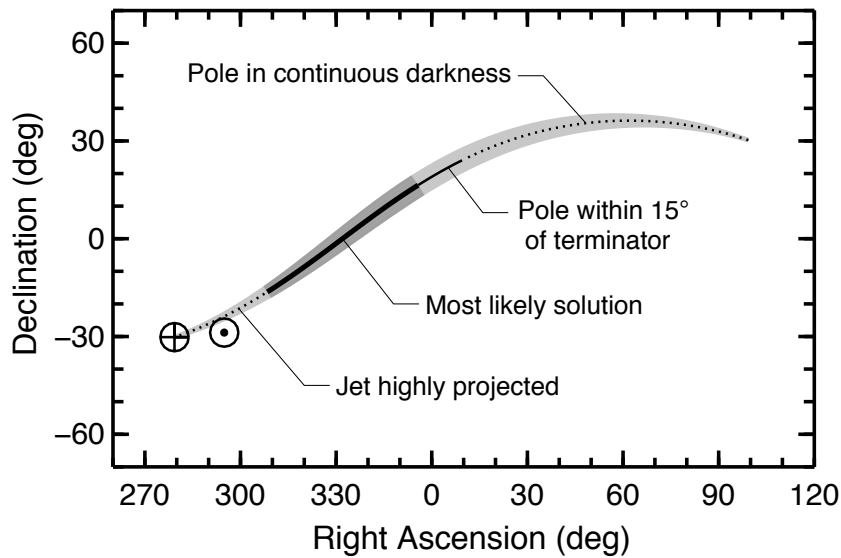

Fig. 4. Plot showing the family of solutions for the pole orientation, as derived from the position angle of the jet. The full extent of the curve delineates the half-great circle of the jet projected on the sky, with additional restrictions imposed by the jet's projection effects and angle with respect to the Sun. The Earth and Sun symbols denote the positions of those bodies as seen from the comet, and the shaded areas define the pole uncertainty.

*3.5 Dust particle size and production rate*

The general coma morphology of C/ISON is consistent with a classic fountain model, with dust grains being turned back into a tail by radiation pressure at a characteristic standoff distance (Grun and Jessberger 1990; Baum et al. 1992). Using this model we derived a characteristic grain size of ~0.5 μm for the observed standoff distance of ~6000 km, assuming grain density of 1 g/cm$^3$ moving at a velocity ~10% of the typical gas velocity $v_g = 0.85 \times r_H^{-0.5}$ km/s, where $r_H$ is the heliocentric distance in AU (Combi 1989). Although this is only a crude analysis, the result suggests that the coma consists of particles in the micron size range traveling at tens of m/s. Adopting a particle size distribution that is rich in micron-sized grains, $Q(a) \sim a^{-3}$, where $Q(a)$ is the production rate of particles with radius $a$, and correcting for the phase angle dependence of $A(\theta)f\rho$, we derive a dust production rate of $\dot{m} \approx 0.14$ kg/s following Fink and Rubin (2012), uncertain by a factor of several depending on the actual particle size distribution, grain scattering function, etc.

**4. Discussion**

Early observations of C/ISON are important to provide the baseline knowledge for detecting and interpreting the later, potentially dramatic evolutionary changes that will occur on the comet. Our *HST* observations of C/ISON help paint the first high spatial resolution, high sensitivity portrait of this comet. The long-term monitoring programs of C/ISON from the ground and space platforms (e.g., Meech et al. 2013; Bodewits et al. 2013) reveal the evolutionary path of the comet, placing the detailed snapshot studies (including Hines et al. 2013, A'Hearn et al. 2013, and Lisse et al. 2013) in perspective.

Given the sungrazing trajectory of C/ISON, the illumination geometry of the comet remains almost unchanged until the last week before perihelion on November 28, 2013. The nucleus' high obliquity from our pole analysis indicates that the surface near the anti-sunward pole is sheltered from sunlight until the comet is within Mercury's orbit (assuming the spin axis remains stable; Samarasinha and Mueller 2013). Due to the generally low thermal inertia of cometary nuclei (e.g., Groussin et al. 2007; 2013), the sunward side of C/ISON's nucleus will be continuously baked, outgassing and evolving, while the anti-sunward side should remain cold and comparatively pristine.

The nearly constant illumination geometry during the approach to the Sun, combined with the high obliquity, has other interesting implications. First, it means that any observed changes in the comet, such as activity level, volatile abundance, coma morphology, etc., must be evolutionary for the same part of the surface, rather than due to seasonal variations. Therefore, pre-perihelion observations of the coma properties of C/ISON can be directly compared to our *HST* studies to search for and study the possible evolutionary effect. Second, because of the constant insolation geometry, the jet observed by *HST* will likely remain active with unchanged morphology before perihelion. Future ground-based observations will be able to observe the jet from different viewpoints to better constrain its orientation, and therefore the rotational pole. However, this conclusion also suggests that it is unlikely that the polar jet will develop diurnal modulation of activity that can be observed as a periodic brightness variation. Any determination of a rotation period will have to rely on the morphology of possibly newly activated jets at lower-latitudes. Third, in the 1-2 weeks bracketing perihelion, a significant part of the original surface will be exposed to strong sunlight for the first time. Since the original surface is formed and possibly altered by galactic cosmic rays when the comet was stored in the OC for billions of years (Stern 1990), the newly illuminated surface could still retain a much higher abundance of very volatile materials, such as CO and/or $CO_2$. The sudden exposure to extremely strong sunlight could trigger enormous outbursts. It will be interesting to see if the first exposure of the nearly pristine surface can be detected by observations near perihelion.

This research is supported by NASA through Grant HST-GO-13198 from the Space Telescope Science Institute.

**References:**
A'Hearn, M.F., et al., 2011, Science 332, 1396.
A'Hearn, M.F., Schleicher, D.G., Feldman, P.D., Millis, R.L., Thompson, D.T., 1984, Astron. J., 89, 579.
A'Hearn, M.F., et al., 2013, 45$^{th}$ DPS Meeting, Abstract #407.04.
ASTM E490, 2006, doi: 10.1520/E0490-00AR06.
Baum, W.A., Kreidl, T.J., & Schleicher, D.G., 1992, Astron. J. 104, 1216.
Beer, E.H., Podolak, M., & Prialnik, D., 2006, Icarus 180, 473.
Bodewits, D., Farnham, T., & A'Hearn, M.F., 2013, CBET #3608.
Clark, R.N., & Lucey, P.G., 1984, J. Geophys. Res. 89, 6341.
Combi, M.R., 1989, Icarus 81, 41.
Combi, M.R., et al., 2013, Icarus 225, 740.
Downs, C., et al., 2013, Science 340, 1196.
Dressel, L., 2012. Wide Field Camera 3 Instrument Handbook, Version 5.0. Baltimore: STScI.
Farnham, T.L., et al., 2013, 45$^{th}$ DPS Meeting, Abstract #407.08.
Feldman, P.D., Cochran, A.L., & Combi, M.R., 2004, in Festou, M., Keller, H.U., Weaver, H.A. (Eds.), Comets II, Univ. Arizona Press, pp. 425.
Fink, U., & Rubin, M., 2012, Icarus 221, 721.
Gonzaga, S., Hack, W., Fruchter, A., & Mack, J., (Eds.), 2002, The DrizzlePac Handbook. (Baltimore, STScI)
Groussin, O., et al., 2007, Icarus 187, 16.
Groussin, O., et al., 2013, Icarus 222, 580.
Grun, E., & Jessberger, E.K., 1990, in: Huebner, W.F. (Ed.), Physics and Chemistry of Comets, Springer-Verlag, New York, pp. 113.
Gustafson, Bo Å. S., 1994, Annu. Rev. Earth Planet. Sci. 22, 553.
Hadamcik, E., & Levasseur-Regourd, A.C., 2009, Planet. Space Sci., 57, 1118.
Hansen, G.B., & McCord, T.B., 2004, J. Geophys. Res. 109, E01012.
Hines, D.C., et al., 2013, Astrophys. J. Lett., submitted.
Jewitt, D., & Meech, K.J., 1986, Astrophys. J., 310, 937.
Kelley, M.S., et al., 2013, Icarus 222, 634.
Knight, M.M., et al., 2013, 45$^{th}$ DPS Meeting, Abstract #407.01.

Kolokolova, L., Jockers, K., Gustafson, B.A.S., & Luchtenberg, G., 2001, J. Geophys. Res. 106, 10113.
Kolokolova, L., Lara, L.M., Schulz, R., Stüwe, J.A., & Tozzi, G.P., 2003, Journal of Quantitative Spectroscopy & Radiative Transfer 79-80, 861.
Köhler, M., Minato, T., Kimura, H., & Mann, I., 2007, Adv. in Space Res. 40, 266.
Li, A., & Greenberg, J.M., 1997, Astron. Astrophys. 323, 566.
Lisse, C.M., et al., 2013, CBET #3598.
Lowry, S.C., Fitzsimmons, A., Cartwright, I.M., & Williams, I.P., 1999, Astron. Astrophys., 349, 649.
Meech, K.J., Svoren, J., 2004, in: Festou, M., Keller, H.U., Weaver, H.A. (Eds.), Comets II, Univ. Arizona Press, pp. 317.
Meech, K.J., et al., 2013, Astrophys. J. Lett., submitted.
Mukai, T. 1989, in Evolution of Interstellar Dust and Related Topics, ed. A. Bonetti, J. M. Greenberg, & S. Aiello (Amsterdam: North-Holland), 397
Nakamura, R., Kitada, Y., & Mukai, T., 1994, Planet. Space Sci. 42, 721.
Nakamura, R., & Hidaka, Y., 1998, Astron. Astrophys. 340, 329.
Nevski, V., & Novichonok, A., 2012, CBET #3238.
Paganini, L., et al., 2012, Astrophys. J. Lett. 748, 13.
Rajan, A., et al., 2010, WFC3 Data Handbook, Version 2.1. Baltimore: STScI.
Samarasinha, N.H., & Mueller, B.E., 2013, Astrophys. J. Lett., accepted.
Schleicher, D., 2013a, IAU Circular #9254.
Schleicher, D., 2013b, IAU Circular #9257.
Schrijver, C.J., et al., 2012, Science 335, 324-328.
Stern, S.A., 1990, Icarus 84, 447.
Weiler, M., Rauer, H., Knollenberg, J., Jorda, L., & Helbert, J., 2003, Astron. Astrophys., 403, 313.
Yang, B., 2013, CBET #3622.